%% file: document.tex
\documentclass[10pt,twocolumn,aps,pra,superscriptaddress,longbibliography,nofootinbib,floatfix,
]{revtex4-2}

\usepackage{times}
\usepackage{graphicx}
\usepackage{dcolumn}
\usepackage{bm}
\usepackage{amsmath}
\usepackage{amssymb}
\usepackage{hyperref}
\hypersetup{colorlinks=true,linkcolor=blue,citecolor=blue,urlcolor=blue}
\usepackage{mathtools}
\usepackage{upgreek}

\begin{document}

\title{Coherence based on positive operator-valued measures for standard and concatenated quantum state discrimination with inconclusive results}

\author{L. F. Melo}%
\email{felipe-melo@ufmg.br}
\affiliation{Departamento de F\'isica, Universidade Federal de Minas Gerais, Belo Horizonte, MG, Brazil, 31270-901}
\affiliation{QuIIN - Quantum Industrial Innovation, EMBRAPII CIMATEC Competence Center in Quantum Tecnologies, SENAI CIMATEC, Av. Orlando Gomes 1845, Salvador, BA, Brazil}

\author{O. Jim\'enez}
\affiliation{Centro Multidisciplinario de Física, Facultad de Ciencias, Universidad Mayor, Camino La Pir\'amide 5750, Huechuraba, Santiago, Chile}

\author{L. Neves}
\affiliation{Departamento de F\'isica, Universidade Federal de Minas Gerais, Belo Horizonte, MG, Brazil}
 
\date{\today}

\begin{abstract}

The optimal measurement that discriminates nonorthogonal quantum states with fixed rates of inconclusive outcomes (FRIO) can be decomposed into an assisted  separation of the inputs, yielding conclusive and inconclusive outputs, followed by a minimum-error (ME) measurement for the conclusive ones (standard FRIO) or both ones (concatenated FRIO). The implementation of these measurements is underpinned by quantum resources, and here we investigate coherence based on positive operator-valued measures (POVMs) as a resource for both strategies in discriminating equally probable symmetric states of arbitrary dimension. First, we show that the POVM coherence in the assisted separation stage decomposes into the coherence of the ancillary state and the quantum discord between the system and the ancilla, evidencing coherence as a more elementary resource than quantum correlations. Next, it is demonstrated that the POVM coherence for standard and concatenated FRIO decomposes into the POVM coherence measures for state separation and ME measurement, weighted by the probabilities of occurrence of each event. Due to the ME discrimination of inconclusive states, the coherence required for the concatenated scheme is shown to be greater than that of the standard one. We discuss other general aspects of our results by characterizing the POVM coherence in the discrimination of qutrit states, with respect to the distinguishability of the inputs and the inconclusive rate. Finally, by exploiting POVM-based coherence as a quantifier of cryptographic randomness gain, we discuss the standard and concatenated FRIO strategies from the perspective of generating random bits that are secret to an eavesdropper. 
\end{abstract}

\maketitle

\section{Introduction} \label{sec:Intro}

Nonorthogonal quantum states cannot be perfectly distinguished due to the laws of quantum theory \cite{BarnettBook}. This limitation gave rise to intense research into optimal measurement strategies for discriminating such states \cite{Chefles00,Barnett09,Bergou10,Bae15}, which later had a strong impact on quantum information and quantum communication \cite{Bennett92,Phoenix00,Banaszek00,Renes04,Brassard04,Pati05,Brask17}. Two strategies stand out as extreme possibilities for this task: the first, minimum error (ME) discrimination, is a deterministic procedure that always infers the state from the measurement outcome, minimizing the average probability of incorrect inferences \cite{HelstromBook,Holevo73,Yuen75}. The second, optimal unambiguous discrimination (UD), is a probabilistic scheme that allows error-free identifications of linearly independent states with a minimum rate of inconclusive results \cite{Ivanovic87,Dieks88,Peres88}. We note that for linearly dependent states, the optimal maximum confidence (MC) strategy generalizes UD, identifying states with the maximum possible confidence, though it cannot be entirely error free \cite{Croke06}. 

It is now known that these fundamental strategies are, in fact, particular cases of a more general scheme called optimal discrimination with fixed rates of inconclusive outcomes (FRIO) \cite{Chefles98-3,Zhang99,Bagan12}. When the inconclusive rate is zero, the optimal FRIO reduces to ME. When this rate is fixed at a critical value, the strategy reduces to optimal UD (or MC, if the confidences are the same for all states \cite{Herzog12}). This strategy provides more confidence in discrimination than ME, with inconclusive rates below the critical value. 

In general, the optimal measurements for quantum state discrimination are based on positive operator-valued measures (POVMs). The POVM for optimal FRIO can be implemented in an assisted manner, by coupling the main system to an auxiliary one (ancilla) and measuring both. This approach decomposes the measurement into a state separation stage that probabilistically changes the distinguishability of the inputs, yielding conclusive (more distinguishable) and inconclusive (less distinguishable) outputs, followed by a ME measurement for the conclusive ones \cite{Chefles98,Nakahira12,Prosser16,Melo23}. We shall refer to this procedure as standard FRIO. In contrast, the procedure in which the inconclusive output states are also discriminated with ME will be called concatenated FRIO. Concatenated discrimination strategies are especially important for high-dimensional states where, in general, inconclusive outputs will still carry useful information about the inputs \cite{Chefles98-1,Peres98,Sun01,Roa11,Jimenez11,Zhang14}. Recovering the information that is discarded in the standard case may improve protocols like teleportation \cite{Neves12}, entanglement swapping \cite{Prosser14}, dense coding, and quantum key distribution \cite{Kogler17}. Recently, the concatenation between optimal UD and ME was experimentally demonstrated \cite{Prosser21}.

The implementation of quantum state discrimination, like many other quantum information tasks, is underpinned by quantum resources, and recent efforts have been directed towards understanding and unraveling these resources. Early studies focused on the role of quantum correlations for standard UD of two \cite{Roa11-2,Zhang13} or more \cite{Xu14,Li12} states, ME \cite{Jimenez19}, and FRIO \cite{Jimenez21} of two states, showing that quantum discord rather than entanglement is the required resource for the protocol. Recent studies, on the other hand, have explored the role of quantum coherence in standard UD of two states \cite{Namkung20,Kim21}, given that it is a more fundamental resource than quantum correlations \cite{Yao15}. However, these works relied on the standard notion of coherence, which is based on projective measurements \cite{Baumgratz14}. A more suitable approach for quantum state discrimination is the generalized resource theory of coherence developed by Bischof \textit{et al.}\ \cite{Bischof19}. In this framework, coherence is the resource needed to implement a POVM on a given state in an extended Hilbert space, which is the operational basis for both standard and concatenated FRIO.

Here, we investigate coherence based on POVMs as a resource for standard and concatenated FRIO discrimination between $N$ equally probable symmetric states of arbitrary dimension $n$, where $N\geqslant n$. First, we show that the POVM coherence in the assisted separation stage decomposes into the coherence of the ancillary state and the quantum discord between the system and the ancilla, evidencing coherence as a more elementary resource than quantum correlations. Next, it is demonstrated that the POVM coherence for standard and concatenated FRIO decomposes into the POVM coherence measures for state separation and ME measurement, weighted by the probabilities of occurrence of each event. Due to the ME discrimination of inconclusive states, the coherence required for the concatenated scheme is shown to be greater than that of the standard one. We discuss other general aspects of our results by characterizing the POVM coherence in the discrimination of qutrit states, with respect to the distinguishability of the inputs and the inconclusive rate. Finally, by exploiting POVM-based coherence as a quantifier of private randomness, we discuss the standard and concatenated FRIO strategies from the perspective of generating random bits that are secret to an eavesdropper---a relevant topic in quantum random number generation \cite{Brask17,Yuan15,Yuan19, Bischof21, Ma19} and quantum cryptography \cite{Ma19-2}.

This paper is structured as follows: In Sec.~\ref{sec:background}, we provide a theoretical background for the subsequent discussion. In Sec~\ref{sec:coherence}, we derive the POVM coherence for standard and concatenated FRIO discrimination, illustrating our results with examples. In Sec.~\ref{sec:discussion}, we discuss these results in light of the operational meaning of POVM coherence. Finally, Sec.~\ref{sec:conclusion} concludes the paper.


\section{Theoretical background}
\label{sec:background}

\subsection{Coherence based on positive operator-valued measures}

Coherence is known to be an intrinsic property of quantum states and an essential ingredient for many quantum phenomena \cite{Streltsov17}. Among the different measures to quantify it \cite{Baumgratz14,Yuan15}, the relative entropy of coherence for a state $\hat{\rho}$ is given by
\begin{equation}  \label{eq:Cr}
C_r(\hat{\rho})=S(\hat{\rho}_\textrm{diag})-S(\hat{\rho}),
\end{equation}
where $S(\hat{\varrho})=-\mathrm{Tr}(\hat{\varrho}\log_2\hat{\varrho})$ is the von Neumann entropy and $\hat{\rho}_\textrm{diag}$ is the state obtained from $\hat{\rho}$ by suppressing its off-diagonal elements. Therefore, diagonal states have zero coherence and are incoherent with respect to a fixed basis. 

The standard notion of coherence is connected with projective measurements in the following sense: given an orthonormal basis $\{|i\rangle\}$ of an $n$-dimensional Hilbert space $\mathcal{H}$, we have $\hat{\rho}=\sum_{i,j=0}^{n-1}\rho_{ij}|i\rangle\langle j|\Rightarrow\hat{\rho}_\textrm{diag}=\sum_{i=0}^{n-1}\hat{\pi}_i\hat{\rho}\hat{\pi}_i$,
where $\hat{\pi}_i=|i\rangle\langle i|$. This means that incoherent states can be seen as arising from a projective measurement on $\hat{\rho}$ in the basis $\{|i\rangle\}$, and coherence, quantified by $C_r(\hat{\rho})$, as the resource required to implement such a measurement. 

Recently, Bischoff \textit{et al.}\ \cite{Bischof19} extended the notion of coherence to encompass POVMs. An $N$-outcome POVM on $\mathcal{H}$ is a set $\bm{\Pi}=\{\hat{\Pi}_i\}_{i=0}^{N-1}$ of positive semidefinite operators, which satisfy $\sum_i\hat{\Pi}_i=\hat{I}$. If $\{\hat{A}_i\}$ denotes a set of detection operators of $\bm{\Pi}$, we have $\hat{\Pi}_i=\hat{A}_i^\dag\hat{A}_i$. By measuring $\hat{\rho}$, the probability to obtain the $i$th outcome is given by $p_i=\mathrm{Tr}(\hat{A}_i\hat{\rho}\hat{A}_i^\dag)$ and the associated postmeasurement state will be $\hat{\rho}_i=\hat{A}_i\hat{\rho}\hat{A}_i^\dag/p_i$. In Ref.~\cite{Bischof19}, the authors show that the coherence resource required to implement this measurement can be quantified by the relative entropy of POVM-based coherence, given by
\begin{align}
    \mathcal{C}_\textrm{rel}(\hat{\rho},\mathbf{\Pi}) = H\left(\lbrace p_i \rbrace\right) + \sum_{i=0}^{N-1} p_i S(\hat{\rho}_i) - S(\hat{\rho}), \label{eq:POVMcoherence}
\end{align}
where $H\left(\lbrace p_i \rbrace\right) = -\sum_i p_i\log_2 p_i$ is the Shannon entropy of the probability distribution $\{p_i\}$. If $\bm{\Pi}$ is a projective measurement, i.e., $\{\hat{A}_i=\hat{\pi}_i\}$, then $\mathcal{C}_\textrm{rel}(\hat{\rho},\mathbf{\Pi})$ reduces to the standard relative entropy of coherence $C_r(\hat{\rho})$. 


\subsection{Quantification of quantum, classical and total correlations in a quantum state}

Consider the case where identical copies of a bipartite system in the state $\hat{\rho}_{da} \in \mathcal{H}_d\otimes \mathcal{H}_a$ are shared between two parties.  In the many copies scenario, the total amount of correlations is given by the quantum mutual information \cite{Groisman04}:
\begin{equation}        \label{eq:mutual}
I(\hat{\rho}_{da}) = S(\hat{\rho}_d) + S(\hat{\rho}_a) - S(\hat{\rho}_{da}),
\end{equation}
where $\hat{\rho}_d=\mathrm{Tr}_a\hat{\rho}_{da}$ and $\hat{\rho}_a=\mathrm{Tr}_d \hat{\rho}_{da}$ are the reduced density matrices for each partition. Now, suppose that for each copy a projective measurement $\lbrace \hat{\pi}^a_i\rbrace$ is implemented on subsystem ``$a$''. The gain of information about ``$d$'' after measuring ``$a$'' is related to the classical correlations from the perspective of the latter subsystem, which are quantified by
\begin{align}   \label{eq:classical}    
J(d |\lbrace \hat{\pi}^a_i \rbrace) = S(\hat{\rho}_d) - \sum_i q_i S(\hat{\rho}^i_{d|a}),
\end{align}
where $q_i = \mathrm{Tr}[(\hat{I}_d\otimes \hat{\pi}^a_i)\hat{\rho}_{da} (\hat{I}_d\otimes \hat{\pi}^a_i)]$ and $\hat{\rho}^i_{d|a} = \mathrm{Tr}_a[(\hat{I}_d\otimes \hat{\pi}^a_i)\hat{\rho}_{da} (\hat{I}_d\otimes \hat{\pi}^a_i)]/q_i$.
The quantum portion of correlations from the perspective of ``$a$'' is measured by the quantum discord, defined as
\begin{align}        
    D(d|a)=I(\hat{\rho}_{da} )-\max_{\{\hat{\pi}_i^a\}}~J (d |\lbrace \hat{\pi}^a_i \rbrace),  \label{eq:discord}
\end{align}
where the maximization of $J$ is over all rank-1 projective measurements on that subsystem. 


\subsection{Optimal discrimination of equiprobable symmetric states with fixed rates of inconclusive outcomes}
\label{subsec:FRIO}

\subsubsection{Parametric separation of equiprobable symmetric states}

Throughout this paper we address the problem of discriminating between $N$ equiprobable symmetric pure states spanning an $n$-dimensional Hilbert space, $\mathcal{H}_d$. These states provide closed-form analytical solutions for the most fundamental state discrimination strategies \cite{Ban97,Chefles98-2,Herzog12,Jimenez11} and are important in many quantum information protocols \cite{Roa03,Renes04,Delgado05,Jimenez10-2,Neves12,Prosser14,Kogler17}. For $j=0,\dots,N-1$, they can be written as
\begin{align}    
    |\alpha_j\rangle &= \sum_{k=0}^{N-1}a_ky_k \omega^{jk}|k\rangle, \label{eq:symmetric} 
\end{align}
where $\omega=\exp(2\pi i/N)$, $\lbrace |k\rangle\rbrace_{k=0}^{N-1}$ is an orthonormal basis spanning an $N$-dimensional Hilbert space, $\{a_k\}$ are nonnegative real coefficients satisfying $\sum_{k}a_k^2=1$, and $y_k\equiv 1-\delta_{0,a_k}$ is a binary parameter such that $n=\dim(\mathcal{H}_d)=\sum_{k=0}^{N-1}y_k\leqslant N$. Thus, the symmetric states will be linearly independent (dependent) if $n=N$ ($n<N$). 

In Ref.~\cite{Prosser16}, the authors demonstrated an optimal scheme to transform an input set $\{|\alpha_j\rangle\}$ into another set of more distinguishable states, with the maximum probability of success. This procedure, known as quantum state separation \cite{Chefles98,Feng05,Dunjko12}, is key to probabilistic discrimination strategies like FRIO and will be summarized here. First, one extends $\mathcal{H}_d$ by attaching an auxiliary system in a two-dimensional Hilbert space, $\mathcal{H}_a$. Next, both systems are coupled through a unitary operation acting on $\mathcal{H}_d\otimes\mathcal{H}_a$ as
\begin{align}          
    \hat{\mathcal{U}}(\xi) |\alpha_j\rangle_d |1\rangle_a &= \sqrt{P(\xi)}|\beta_j(\xi)\rangle_d |1\rangle_a + \sqrt{Q(\xi)} |\tilde{\beta}_j\rangle_d |0\rangle_a. \label{eq:unitary}
\end{align}
In this expression, $\xi\in[0,1]$ and $\lbrace|0\rangle_a, |1\rangle_a\rbrace$ is an orthonormal basis for $\mathcal{H}_a$; the states $|\beta_j(\xi)\rangle_d$ and $|\tilde{\beta}_j\rangle_d$ are given by
\begin{subequations} 
\begin{align}
   |\beta_j(\xi)\rangle  &= \sum_{k=0}^{N-1} \underbrace{\sqrt{(1-\xi)a_k^2 + \frac{y_k\xi}{n}}}_{\displaystyle b_k(\xi)}\omega^{jk}|k\rangle , \label{eq:beta}\\
    |\tilde{\beta}_j\rangle &= \sum_{k=0}^{N-1}\underbrace{\sqrt{\frac{a_k^2-a_\textrm{min}^2y_k}{1-na_\textrm{min}^2}}}_{\displaystyle \tilde{b}_k}\omega^{jk}|k\rangle, \label{eq:betatilde}
\end{align}
\end{subequations}
respectively, and 
\begin{equation}  \label{eq:successprobability} 
    P(\xi) = \frac{na_\textrm{min}^2}{(1-\xi)na_\textrm{min}^2 + \xi} = 1 - Q(\xi)
\end{equation}
are the optimal probabilities of success $(P)$ and failure $(Q)$,
where $a_\textrm{min}=\min_k\{a_k\}\neq 0$.\footnote{The coefficient $a_\mathrm{min}$ minimizes the probability of failure while ensuring the positive semi-definiteness of the POVM element associated with this result (e.g., see Refs.~\cite{Jimenez11,Prosser16}).} Note that for parallel inputs ($a_k=\delta_{kl}$) and $\xi>0$, we assume $a_\textrm{min}=0$, which ensures that $P(\xi>0)=0$.
Finally, after this coupling, the ancilla is measured in the basis $\lbrace|0\rangle_a, |1\rangle_a\rbrace$: the projection onto $|1\rangle_a$ occurs with the maximum probability of success $P(\xi)$ and leads to the desired transformation $|\alpha_j\rangle\rightarrow|\beta_j(\xi)\rangle$, otherwise the process fails with probability $Q(\xi)$ and $|\alpha_j\rangle\rightarrow|\tilde{\beta}_j\rangle$.

The parameter $\xi$ sets the degree of separation: for $\xi=0$ there is no change of the inputs, i.e., $|\beta_j(0)\rangle=|\alpha_j\rangle\,\forall\,j$; in the range $0<\xi\leqslant 1$, the distinguishability increases monotonically with $\xi$, and for $\xi=1$, the successfully transformed states become maximally distinguishable:  
\begin{equation}    \label{eq:uj}
|\beta_j(1)\rangle\equiv|u_j\rangle=\frac{1}{\sqrt{n}}\sum_{k=0}^{N-1}y_k\omega^{jk}|k\rangle
\end{equation}
with probability $P(1)=na_\textrm{min}^2$. For instance, if $N=n$, the states $\{|u_j\rangle\}$ will be orthogonal. 

The states $\{|\tilde{\beta}_j\rangle\}$ resulting from a failure in the process are independent of $\xi$ [see Eq.~(\ref{eq:betatilde})] and span a ($n-\mu$)-dimensional space, where $\mu\equiv\mu(a_\textrm{min})$ is the multiplicity of $a_\textrm{min}$. Therefore, when the separation fails, with the minimum probability $Q(\xi)$, the output states become less distinguishable than the inputs, but will still carry information about them if $n-\mu>1$; this is the appropriate scenario to concatenate discrimination strategies \cite{Sun01, Roa11, Zhang14,Jimenez11,Prosser21}.


\subsubsection{Standard and concatenated optimal FRIO measurement for symmetric states}

The optimal FRIO strategy can be decomposed into two steps \cite{Nakahira12,Prosser16}: first, one implements state separation, where the pre-established value of $\xi$ will fix the rate of inconclusive outcomes arising from failed events. Next, ME discrimination is applied on the states emerging from the successful separation (standard FRIO) or both successful and failed separation (concatenated FRIO).

The quantum state separation process outlined above can be described by the two-outcome POVM $\bm{\Pi}_\textsc{sep}=\{\hat{A}_s^\dag(\xi)\hat{A}_s(\xi), \hat{A}_f^\dag(\xi)\hat{A}_f(\xi)\}$ on $\mathcal{H}_d$, where 
\begin{subequations}
\begin{align}
    \hat{A}_s(\xi) & ={}_a\langle 1|\hat{\mathcal{U}}(\xi)|1\rangle_a, \label{eq:POVMsepS}\\
    \hat{A}_f(\xi) &= {}_a \langle 0| \hat{\mathcal{U}}(\xi)|1\rangle_a  \label{eq:POVMsepF}
\end{align}
\end{subequations}
are the detection operators associated with the success and failure outcomes, respectively, and $\hat{\mathcal{U}}(\xi)$ is the unitary given by Eq.~(\ref{eq:unitary}). On the other hand, the optimized measurement that discriminates between $N$ equally likely symmetric states with the minimum average probability of error is given by the $N$-outcome POVM $\bm{\Pi}_\textsc{me}=\{\hat{\Pi}^{\textsc{me}}_j\}_{j=0}^{N-1}$, where \cite{Ban97,Jimenez11}
\begin{align}
    \hat{\Pi}^{\textsc{me}}_j=\frac{n}{N} |u_j\rangle\langle u_j|, \label{eq:MEPOVM}
\end{align}
and $|u_j\rangle$ is given by Eq.~(\ref{eq:uj}); it applies to both successful ($|\beta_j(\xi)\rangle$) and failure ($|\tilde{\beta}_j\rangle$) states emerging from the separation step, as they are also equiprobable and symmetric [see Eqs.~(\ref{eq:beta}) and (\ref{eq:betatilde})].

In the standard FRIO discrimination, where failed events are discarded as inconclusive results, the optimal measurement is given by an $(N+1)$-outcome POVM $\bm{\Pi}_\textsc{frio}=\{\hat{\Pi}_0(\xi),\ldots,\hat{\Pi}_{N-1}(\xi), \hat{\Pi}^?(\xi)\}$, where $\hat{\Pi}_j(\xi)=\hat{A}_j^\dag(\xi)\hat{A}_j(\xi)$ and $\hat{\Pi}^?(\xi)=\hat{A}^{?\dag}(\xi)\hat{A}^?(\xi)$, with the corresponding detection operators 
\begin{subequations} \label{eq:friopovm}
\begin{align}
    \hat{A}_j(\xi) &= \sqrt{\hat{\Pi}^{\textsc{me}}_j} \hat{A}_s (\xi), \label{eq:friopovmj}\\
    \hat{A}^?(\xi) &= \hat{A}_f (\xi), \label{eq:friopovm?}
\end{align}
\end{subequations}
associated with a conclusive identification of the input state (whether it is correct or not) and an inconclusive answer, respectively. In contrast, the failures are not discarded in the concatenated FRIO, which is given by a $2N$-outcome POVM 
$\bm{\Pi}_\textsc{conc}=\{\hat{\Pi}_0(\xi),\ldots,\hat{\Pi}_{N-1}(\xi), \hat{\Pi}^?_0(\xi),\ldots,\hat{\Pi}^?_{N-1}(\xi)\}$, where $\hat{\Pi}_j(\xi)=\hat{A}_j^\dag(\xi)\hat{A}_j(\xi)$ and $\hat{\Pi}_j^?(\xi)=\hat{A}_j^{?\dag}(\xi)\hat{A}_j^?(\xi)$, with the corresponding detection operators
\begin{subequations}  \label{eq:POVM_conc}
\begin{align}
    \hat{A}_j(\xi) &= \sqrt{\hat{\Pi}^{\textsc{me}}_j} \hat{A}_s(\xi), \label{eq:concAj}\\
    \hat{A}^?_j(\xi) &= \sqrt{\hat{\Pi}^{\textsc{me}}_j} \hat{A}_f(\xi), \label{eq:concAf}
\end{align}
\end{subequations}
both associated with a conclusive identification of the input arising from successful and failed events, respectively.


\section{POVM-based coherence in quantum state discrimination with FRIO} \label{sec:coherence}

With the theoretical framework presented above, we are now able to investigate the role of POVM-based coherence in the FRIO discrimination of equiprobable symmetric states. To avoid cumbersome equations, from now on we will omit the dependence on the separation parameter $\xi$ of all entities that are functions of it.

\subsection{POVM coherence in quantum state separation}\label{subsec:coherenceseparation}

In the state separation process, let $\hat{\rho}$, $\hat{\rho}_s$, and $\hat{\rho}_f$ denote the density matrices describing the input, successful output, and failed output states, respectively. Using Eqs.~(\ref{eq:symmetric}), (\ref{eq:beta}) and (\ref{eq:betatilde}), and the fact that $\omega$ is an $N$th root of unity, thus satisfying $\sum_{j=0}^{N-1}\omega^{j(k-k')}=N\delta_{k,k'}$, these states will be given by
\begin{subequations} \label{eq:rho_dsf}
\begin{align}  
    \hat{\rho}&=\frac{1}{N}\sum_{j=0}^{N-1} |\alpha_j\rangle \langle\alpha_j|= \sum_{k=0}^{N-1} a_k^2 y_k |k\rangle\langle k|, \label{eq:inputrho} \\
    \hat{\rho}_s &=\frac{1}{N}\sum_{j=0}^{N-1} |\beta_j\rangle \langle\beta_j|= \sum_{k=0}^{N-1} b_k^2y_k |k\rangle\langle k|, \label{eq:rhos} \\
    \hat{\rho}_f &=\frac{1}{N}\sum_{j=0}^{N-1} |\tilde{\beta}_j\rangle \langle\tilde{\beta}_j|= \sum_{k=0}^{N-1} \tilde{b}_k^2y_k |k\rangle\langle k|. \label{eq:rhof}
\end{align}
\end{subequations}
Thus, denoting the global system-ancilla state after the unitary coupling (\ref{eq:unitary}) as $\hat{\rho}_{da}$, we have
\begin{align}    \label{eq:bipartite}
\hat{\rho}_{da} = & \; 
\hat{\mathcal{U}}\left(\hat{\rho}\otimes|1\rangle_a\langle 1|\right) \hat{\mathcal{U}}^\dagger\nonumber\\= & \; P \hat{\rho}_s\otimes |1\rangle_a\langle 1| + Q \hat{\rho}_f \otimes |0\rangle_a\langle 0| \nonumber \\
& \text{} +\sqrt{PQ}\sum_{k=0}^{N-1} b_k\tilde{b}_k |k\rangle\langle k|\otimes \hat{\sigma}_x^a,
\end{align}
where $\hat{\sigma}_x^a = |1\rangle_a\langle0| + |0\rangle_a\langle 1|$. It can be shown that the reduced state of the system is preserved, i.e., 
\begin{equation}    \label{eq:rho_d}
\hat{\rho}_d = \mathrm{Tr}_a\hat{\rho}_{da} = P\hat{\rho}_s+Q\hat{\rho}_f = \hat{\rho},
\end{equation} 
while the ancilla state $\hat{\rho}_a = \mathrm{Tr}_d\hat{\rho}_{da}$ is transformed into
\begin{align}
\hat{\rho}_a =\,P|1\rangle_a\langle 1| + Q|0\rangle_a\langle 0| + \sqrt{PQ} \sum_{k=0}^{N-1} b_k \tilde{b}_k \hat{\sigma}_x^a\label{eq:ancilla}.
\end{align}
It can be seen that coherence, in the sense of its standard resource theory \cite{Baumgratz14}, is produced only in the ancilla state. We quantify it by the relative entropy of coherence given by Eq.~(\ref{eq:Cr}), which in the basis $\lbrace |0\rangle_a, |1\rangle_a \rbrace$ yields
\begin{align}     
C_r(\hat{\rho}_a)=H_2(P)-S(\hat{\rho}_a),\label{eq:Crb}
\end{align}
where $H_2(x) = - x\log_2x - (1-x) \log_x(1-x)$ is the binary entropy and $S(\hat{\rho}_a)=-\sum_{j=\pm}\lambda_j\log_2\lambda_j$, with $\lambda_\pm=\frac{1}{2}\lbrace 1\pm\sqrt{1-4PQ[1-(\sum_k b_k\tilde{b}_k)^2]}\rbrace$ denoting the eigenvalues of $\hat{\rho}_a$.

The coherence resource required to implement the POVM for state separation, $\bm{\Pi}_\textsc{sep}$, on the input state $\hat{\rho}$ is not fully described by the coherence produced in the ancillary system given by Eq.~(\ref{eq:Crb}). To see this, we first compute the POVM coherence using Eq.~(\ref{eq:POVMcoherence}), obtaining
\begin{align}  \label{eq:separationcoherence}
    \mathcal{C}_\textrm{rel}(\hat{\rho},\mathbf{\Pi}_{\textsc{sep}}) &= H_2(P) + P S(\hat{\rho}_s) + Q S(\hat{\rho}_f)-S(\hat{\rho}),
\end{align}
where, from Eqs.~(\ref{eq:rho_dsf}), the von Neumman entropies will be simply given by the Shannon entropies $H(\{x_k^2\})$, with $x=a,b,\tilde{b}$. Now, we address the correlations involved in the process: using the fact that $S(\hat{\rho}_\textrm{d})= S(\hat{\rho})$ and the invariance of entropy under unitary transformations, the quantum mutual information between system and ancilla [see Eq.~(\ref{eq:mutual})] will be 
\begin{align}    \label{eq:mutualsep}
I(\hat{\rho}_{da}) = S(\hat{\rho}_a),
\end{align}
showing that the ``cost'' of generating correlations is the loss of purity in the ancilla. As demonstrated in Refs.~\cite{Jimenez19,Jimenez21}, the classical portion of such correlations from the perspective of the ancilla is maximized by the projective measurement in the basis $\lbrace|0\rangle_a,|1\rangle_a\rbrace$; thus, from Eq.~(\ref{eq:classical}) we have
\begin{align}     \label{eq:J}
    \max~J(d|\lbrace \hat{\pi}_i^a\rbrace) = S(\hat{\rho}) -P S(\hat{\rho}_s) - Q S(\hat{\rho}_f).
\end{align}
Finally, using the definition of quantum discord [Eq.~(\ref{eq:discord})] and Eqs.~(\ref{eq:Crb})--(\ref{eq:J}), we can rewrite the POVM coherence (\ref{eq:separationcoherence}) as
\begin{align}  \label{eq:coherencediscord}
    \mathcal{C}_\textrm{rel}(\hat{\rho}, \mathbf{\Pi}_{\textsc{sep}}) = C_r(\hat{\rho}_a)+ D(d|a).
\end{align}
Therefore, the required coherence for implementing the optimal POVM for state separation is built both from the generated coherence in the ancilla and the quantum correlations between system and ancilla. 

\begin{figure*}
\centerline{\includegraphics[width=1\textwidth]{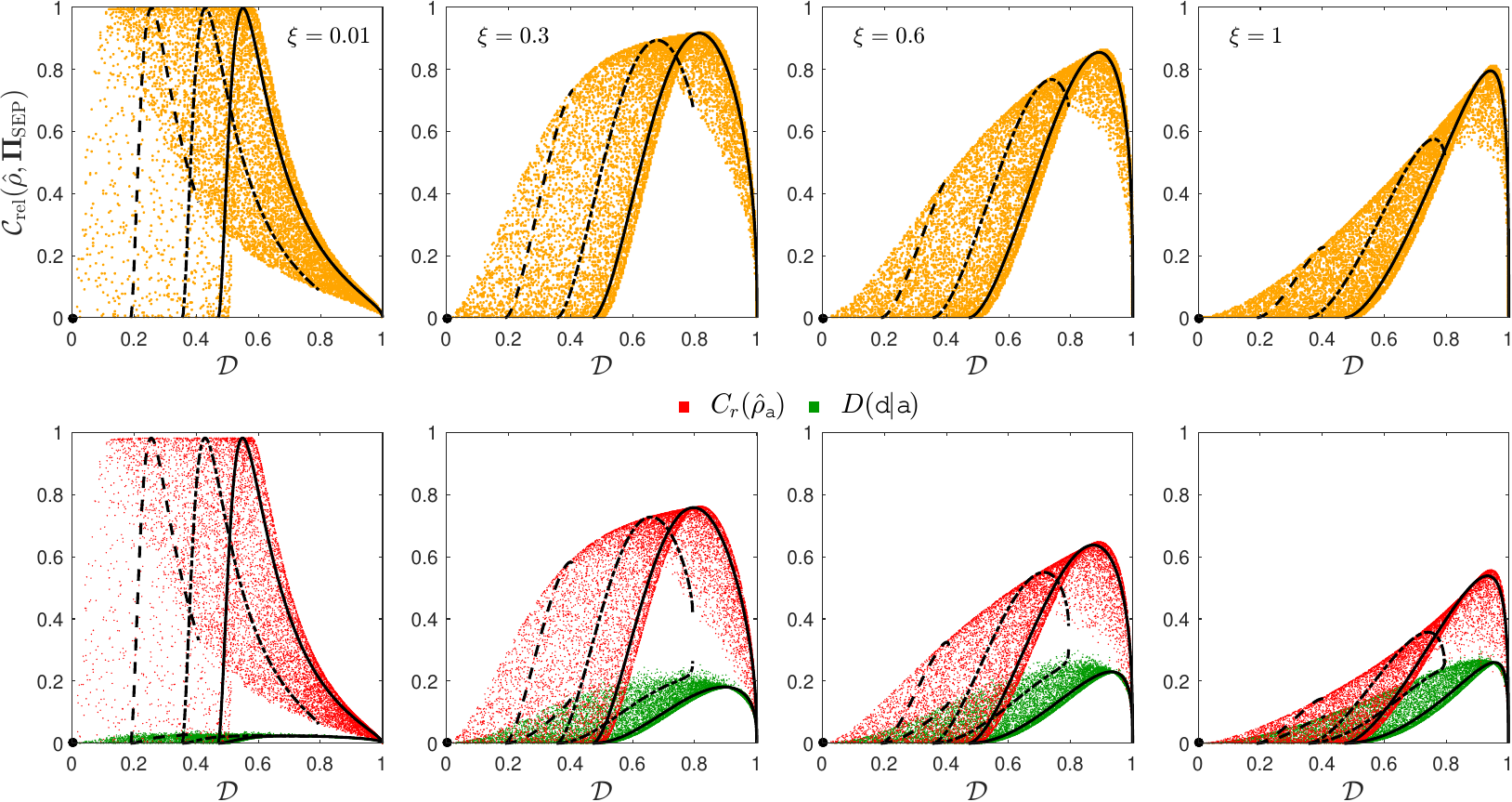}}
\caption{POVM coherence (orange dots), ancilla's coherence (red dots), and quantum discord (green dots) vs distinguishability in the optimal separation of $N=3$ symmetric states of a qutrit. These quantities were computed for $10^4$ random input states and a fixed value of the separation parameter $\xi$ shown in the insets. We also plot them for inputs in the range $a_\textrm{min}\in[0,a_0]$ [see Eq.~(\ref{eq:coefficients})], where $a_0=0$ (circle), $a_0=0.192$ (dashed lines), $a_0=0.385$ (dash-dotted lines), and $a_0=1/\sqrt{3}$ (solid lines).}
\label{fig:separation} 
\end{figure*}

\subsubsection{Example: Separating \texorpdfstring{$N$}{N} symmetric states of a qutrit}

To discuss general aspects of the above result, let us take as an example the separation between $N$ symmetric states of a qutrit ($n=3$). We study the coherence in the process as a function of the distinguishability of the input states $\{|\alpha_j\rangle\}$, which can be quantified by
\cite{Bagan16}
\begin{align}   \label{eq:Dist_ME}
\mathcal{D} &= \frac{n}{n-1}\left(P_\textrm{corr}-\frac{1}{N}\right),
\end{align}
where $P_\textrm{corr}$ is the average probability of correctly identifying them through the ME measurement. Using Eq.~(\ref{eq:MEPOVM}), we have
\begin{align}  \label{eq:Pcorr}
P_\textrm{corr} & = \frac{1}{N}\sum_{j=0}^{N-1}\langle\alpha_j|\hat{\Pi}_j^\textsc{me}|\alpha_j\rangle =\frac{1}{N}\left(\sum_{j=0}^{N-1}a_j\right)^2.
\end{align}
This measure is bounded by $0\leqslant\mathcal{D}\leqslant n/N$, where the lower and upper bounds are attained by parallel and maximally distinguishable input states, respectively.

\begin{figure}[t]
\centerline{\includegraphics[width=1\columnwidth]{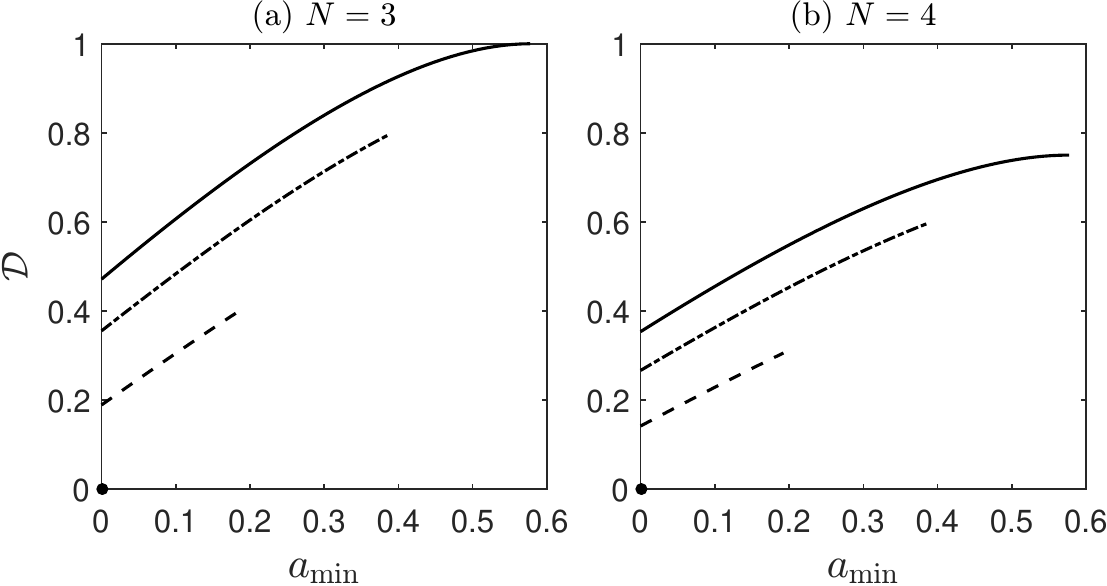}}
  \caption{Distinguishability vs $a_\textrm{min}\in[0,a_0]$ for (a) $N=3$ and (b) $N=4$ symmetric states of a qutrit defined by the coefficients given by Eq.~(\ref{eq:coefficients}), where $a_0=0$ (circle), $a_0=0.192$ (dashed line), $a_0=0.385$ (dash-dotted line), and $a_0=1/\sqrt{3}$ (solid line).}
\label{fig:DistME}
\end{figure}

Considering $N=3$, we generate $10^4$ random input states $\hat{\rho}$ [see Eq.~(\ref{eq:inputrho})], and from each state we compute the three terms of Eq.~(\ref{eq:coherencediscord}) for a fixed value of the separation parameter $\xi$. The results obtained as a function of $\mathcal{D}$ are shown in Fig.~\ref{fig:separation}: in the top row, we plot the POVM coherence for state separation (orange dots), and in the bottom row its components $C_r(\hat{\rho}_a)$ (red dots) and $D(d|a)$ (green dots); from left to right, we have $\xi=0.01,0.3,0.6,$ and $1$.\footnote{Note that the POVM coherence for state separation does not depend on $N$. Thus, for $N>3$ it would have the same behavior shown in Fig.~\ref{fig:separation}, but in a smaller range for distinguishability, i.e., $\mathcal{D}\in[0,n/N]$.}  To assist in the analysis, we also computed these quantities considering input states generated with the coefficients $(a_0,a_1,a_2)$ given by
\begin{subequations}   \label{eq:coefficients}
\begin{align}  
a_0&= \{0,0.192,0.385,1/\sqrt{3}\}, \\
a_1&=a_\textrm{min}\in[0,a_0], \\
a_2&=\sqrt{1-a_0^2-a_1^2},
\end{align}
\end{subequations}
that is, we choose a fixed value for $a_0$ which sets a variable $a_1=a_\textrm{min}$ ($a_2$ is obtained from the normalization condition). The sets of states in each interval $a_\textrm{min}\in [0,a_0]$ are sorted in ascending order with respect to distinguishability, as shown in Fig.~\ref{fig:DistME} for $N=3$ and $4$.

In Fig.~\ref{fig:separation} we plot $\mathcal{C}_\textrm{rel}(\hat{\rho}, \mathbf{\Pi}_{\textsc{sep}})$, $C_r(\hat{\rho}_a)$, and $D(d|a)$ for $N=3$ states in the range $[0,a_0]$, where $a_0=0$ (circle), $a_0=0.192$ (dashed lines), $a_0=0.385$ (dash-dotted lines), and $a_0=1/\sqrt{3}$ (solid lines). For each range, the POVM coherence with respect to the distinguishability of the inputs, according to its components, either presents a monotonic increase\footnote{Our selected sets of states given by Eqs.~(\ref{eq:coefficients}) cannot capture this behavior for $\xi=0.01$.} for low values of $\mathcal{D}$ or exhibits a nonmonotonic behavior otherwise. Later, we shall see that these behaviors influence the coherence for standard and concatenated FRIO measurements and are in sharp contrast with the coherence for ME. Then, we will exploit these results in Sec.~\ref{subsec:Private} to compare the measurement strategies regarding the private randomness gain provided by each one.

Figure~\ref{fig:separation} shows that the main contribution to $\mathcal{C}_\textrm{rel}(\hat{\rho}, \mathbf{\Pi}_{\textsc{sep}})$ comes from the ancilla's coherence. However, it can be shown that the total correlations between system and ancilla increase with the separation degree. Consequently, Eq.~(\ref{eq:mutualsep}) implies that the ancilla's state becomes less pure with the transformation. As a result, the ancilla's coherence [Eq.~(\ref{eq:Crb})] decreases with $\xi$, and its contribution becomes more balanced with the quantum correlations between the two parts. The POVM coherence is zero only at the boundaries of $\mathcal{D}$, where $P=0$ and $1$, so that no further separation is possible in those cases. For $0<\mathcal{D}<n/N$, it can be arbitrarily close to zero if $P\approx 0$ (e.g., for states with arbitrarily small $a_\textrm{min}$) or $P\approx 1$ (e.g., for arbitrarily small $\xi$). In this range, we also observe that the separation of input states with the same distinguishability will require different amounts of POVM coherence, which depend on an intricate relationship between the characteristics of the inputs\footnote{Their coefficients $\{a_j\}$, the value of $a_\textrm{min}$ and its multiplicity $\mu$, and the associated output states $\hat{\rho}_s$ and $\hat{\rho}_f$.} and the desired degree of separation. It can be shown that there are input states with the same $\mathcal{D}$ and quite distinct characteristics, which present opposite behaviors for coherence depending on $\xi$.  

\subsection{POVM coherence in standard FRIO measurement}\label{sec:frio}

The $(N+1)$-outcome POVM for the optimal standard FRIO measurement is built by the detection operators of Eq.~(\ref{eq:friopovm}), and the associated probabilities $\{p_0,\ldots.p_{N-1},p^?\}$ will be given by
\begin{subequations}
\begin{align}
    p_j &= \mathrm{Tr}[\hat{A}_j \hat{\rho}\hat{A}_j^\dag ] = \frac{P}{N}, \label{eq:pjfrio}\\
    p^? &=\mathrm{Tr}[\hat{A}^? \hat{\rho}\hat{A}^{?\dag} ]= Q.   \label{eq:p?}
\end{align}
\end{subequations}
The postmeasurement states for the conclusive and inconclusive outcomes are, respectively, $\hat{\rho}'_j=\hat{A}_j\hat{\rho}\hat{A}_j^\dag/p_j=|u_j\rangle\langle u_j|$ and $\hat{\rho}^?=\hat{A}^?\hat{\rho}\hat{A}^{?\dag}/p^?=\hat{\rho}_f$ [Eq.~(\ref{eq:rhof})], so that $S(\hat{\rho}'_j)=0$ and $S(\hat{\rho}_f)\geqslant 0$. Therefore, from Eq.~(\ref{eq:POVMcoherence}), the required coherence to implement this POVM on the input state $\hat{\rho}$ will be
\begin{align}
   \mathcal{C}_\textrm{rel}(\hat{\rho},\mathbf{\Pi}_{\textsc{frio}})= H_2(P) + P \log_2 N + QS(\hat{\rho}_f) - S(\hat{\rho}). \label{eq:friocoherence}
\end{align}
We can rewrite this expression in a more instructive form by noting that the coherence required to implement the ME measurement on $\hat{\rho}$ is
\begin{align}     \label{eq:MEcoherence}
    \mathcal{C}_\textrm{rel}(\hat{\rho}, \mathbf{\Pi}_{\textsc{me}}) &= \mathcal{C}_\textrm{rel}(\hat{\rho},\mathbf{\Pi}_{\textsc{frio}})|_{\xi=0} \nonumber\\
    &= \log_2 N - S(\hat{\rho}).
\end{align}
Then, using Eq.~(\ref{eq:separationcoherence}), we obtain
\begin{align}
    \mathcal{C}_\textrm{rel}(\hat{\rho},\mathbf{\Pi}_{\textsc{frio}})  = \mathcal{C}_\textrm{rel}(\hat{\rho},\mathbf{\Pi}_{\textsc{sep}})+ P\mathcal{C}_\textrm{rel}(\hat{\rho}_s,\mathbf{\Pi}_{\textsc{me}}), \label{eq:sumofcoherences}
\end{align}
where $\mathcal{C}_\textrm{rel}(\hat{\rho}_s, \mathbf{\Pi}_{\textsc{me}}) = \log_2 N - S(\hat{\rho}_s)$. Reflecting the two-step nature of its implementation, the consumed coherence in the optimal FRIO discrimination is, thereby, split in the coherences for implementing state separation and the ME measurement on $\hat{\rho}_s$ weighted by the success rate.


\begin{figure}[t]
\centerline{\includegraphics[width=1\columnwidth]{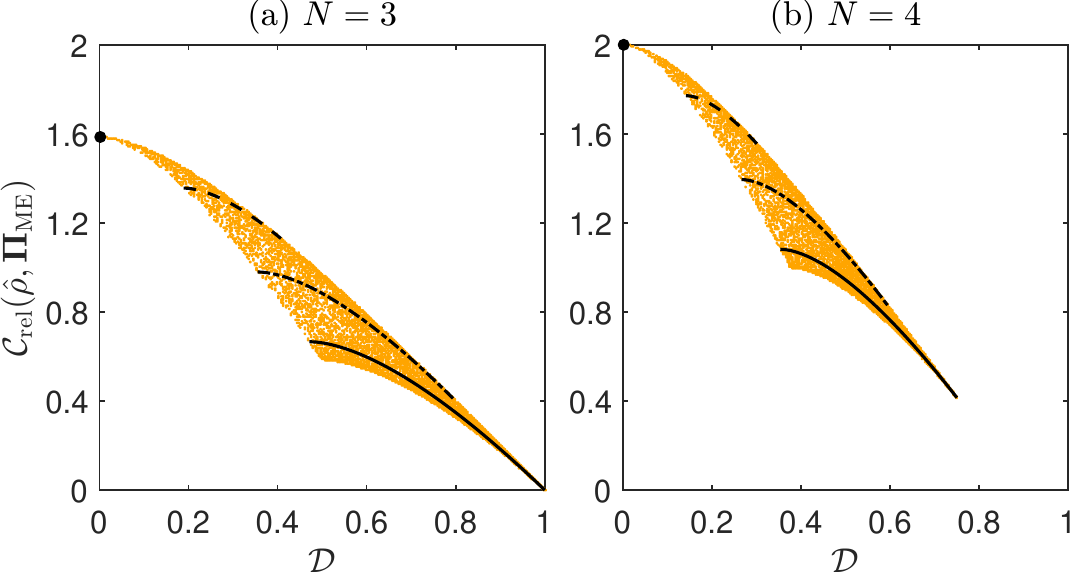}}
\caption{POVM coherence vs distinguishability in the ME discrimination of (a) $N=3$ and (b) $N=4$ symmetric states of a qutrit. The orange dots represent $10^4$ random input states while the plots represent the inputs in the range $a_\textrm{min}\in[0,a_0]$ [see Eq.~(\ref{eq:coefficients})], where $a_0=0$ (circle), $a_0=0.192$ (dashed line), $a_0=0.385$ (dash-dotted line), and $a_0=1/\sqrt{3}$ (solid line).}
\label{fig:POVM_ME}
\end{figure}

\begin{figure*}[t]
\includegraphics[width=\textwidth]{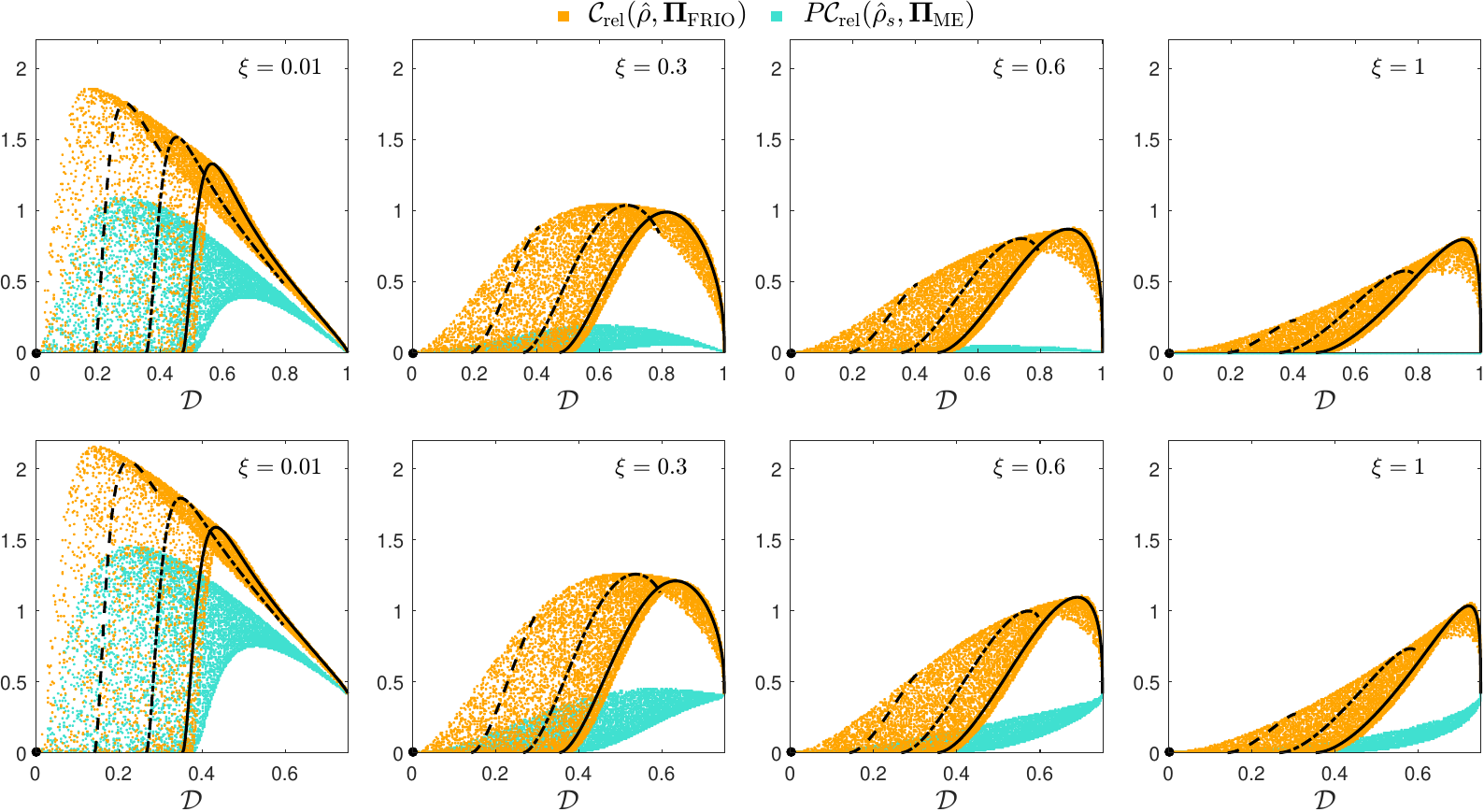} 
\caption{POVM coherence (orange dots) and its component $P\mathcal{C}_\textrm{rel}(\hat{\rho}_s,\bm{\Pi}_\textsc{me})$ (blue dots) vs distinguishability in the standard FRIO discrimination of $N=3$ (top row) and $4$ (bottom row) symmetric states of a qutrit. These quantities were computed for $10^4$ random input states and a fixed value of the separation parameter $\xi$ shown in the insets. We also plot the former for inputs in the range $a_\textrm{min}\in[0,a_0]$ [see Eq.~(\ref{eq:coefficients})], where $a_0=0$ (circle), $a_0=0.192$ (dashed lines), $a_0=0.385$ (dash-dotted lines), and $a_0=1/\sqrt{3}$ (solid lines).}
\label{fig:POVM_FRIO}
\end{figure*}

\subsubsection{Example: Discriminating \texorpdfstring{$N$}{N} symmetric states of a qutrit with standard FRIO}

Returning to the example introduced in Sec.~\ref{subsec:coherenceseparation}, we now study the coherence in the optimal FRIO discrimination between $N$ symmetric states of a qutrit. In what follows, we again use $10^4$ random input states as well as fixed inputs defined by the coefficients in (\ref{eq:coefficients}).

First, considering the ME measurement for $N=3$ and $4$, Fig.~\ref{fig:POVM_ME} shows the corresponding POVM coherence [Eq.~(\ref{eq:MEcoherence})] as a function of distinguishability for the random input states (orange dots); the plots for the fixed inputs (black circle and lines) reveal a monotonic decreasing of $\mathcal{C}_\textrm{rel}(\hat{\rho},\mathbf{\Pi}_{\textsc{me}})$  with respect to $\mathcal{D}$. These results show a complementary relationship between this coherence and distinguishability.\footnote{In fact, the complementary nature between $\mathcal{D}$ given by Eq.~(\ref{eq:Dist_ME}) and coherence quantified by the $l_1$ norm has been found in Ref.~\cite{Bagan16}. Here, as the POVM coherence is quantified by an entropic measure, we should also define an entropic measure of distinguishability to establish a proper complementarity relation between these quantities, but this is beyond the scope of the present paper.} The minimum of $\mathcal{C}_\textrm{rel}(\hat{\rho},\mathbf{\Pi}_{\textsc{me}})$, $\log_2N/n$, is reached for maximally distinguishable inputs, showing that no coherence is consumed in the discrimination of orthogonal states. The maximum, $\log_2N$, is reached for parallel input states, in which case no information is acquired and one obtains the maximum randomness of outcomes (see Sec.~\ref{subsec:Bounds}).

\begin{figure}[t]
\centerline{\includegraphics[width=1\columnwidth]{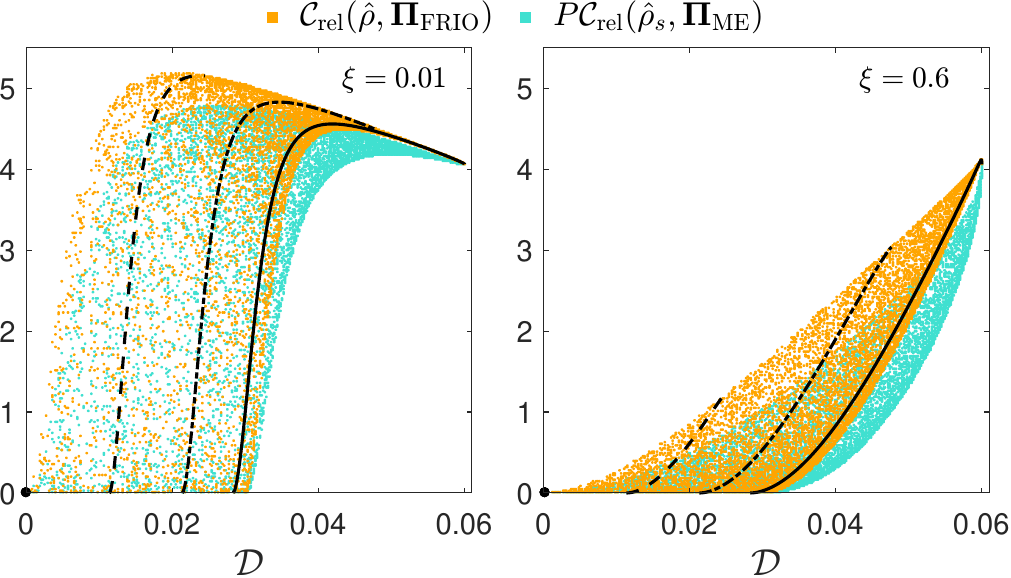}}
  \caption{POVM coherence (orange dots) and its component $P\mathcal{C}_\textrm{rel}(\hat{\rho}_s,\bm{\Pi}_\textsc{me})$ (blue dots) vs distinguishability in the standard FRIO discrimination of $N=50$ symmetric states of a qutrit. These quantities were computed for $10^4$ random input states and a fixed value of the separation parameter $\xi$ shown in the insets. We also plot the former for inputs in the range $a_\textrm{min}\in[0,a_0]$ [see Eq.~(\ref{eq:coefficients})], where $a_0=0$ (circle), $a_0=0.192$ (dashed lines), $a_0=0.385$ (dash-dotted lines), and $a_0=1/\sqrt{3}$ (solid lines).}
\label{fig:CorFRIOlargeN}
\end{figure}

We now address the coherence required for the standard FRIO measurement ($\xi>0$). Using Eq.~(\ref{eq:sumofcoherences}), in Fig.~\ref{fig:POVM_FRIO} we plot $\mathcal{C}_\textrm{rel}(\hat{\rho},\mathbf{\Pi}_{\textsc{frio}})$ (orange dots) and the second term on the right-hand side (blue dots) as a function of $\mathcal{D}$ and $\xi$ for the random input states; we also plot the former for the fixed inputs (black circle and lines). The top (bottom) row corresponds to $N=3$ ($N=4$) states to be discriminated. Note that the contribution from the first term on the right-hand side of (\ref{eq:sumofcoherences}), namely the POVM coherence for state separation, was  discussed earlier (e.g., see Fig.~\ref{fig:separation}). In the linearly independent case (top row), the coherence behavior is mainly dictated by the state separation stage (which is true whatever $n=N$), since the contribution from the second term in (\ref{eq:sumofcoherences}) is relevant only for $\xi\ll 1$. This occurs because the successfully separated states are more distinguishable, so that the required coherence for its ME discrimination, $\mathcal{C}_\textrm{rel}(\hat{\rho}_s, \mathbf{\Pi}_{\textsc{me}})$, decreases with $\xi$. As a consequence, the coherence required to implement optimal UD ($\xi=1$) is fully consumed at the separation stage, i.e.,
\begin{align}
    \mathcal{C}_\textrm{rel}(\hat{\rho},\mathbf{\Pi}_{\textsc{ud}})  = \mathcal{C}_\textrm{rel}(\hat{\rho},\mathbf{\Pi}_{\textsc{sep}})|_{\xi=1}. \label{eq:UD}
\end{align} 
On the other hand, in the linearly dependent case (bottom row), the contribution from the second term in (\ref{eq:sumofcoherences}) is relevant for any $\xi$. In fact, since the coherence from state separation does not depend on $N$, the contribution of $P\mathcal{C}_\textrm{rel}(\hat{\rho}_s,\mathbf{\Pi}_{\textsc{me}})$ becomes dominant as the ratio $N/n$ increases. If $N\gg n$, this contribution dictates the coherence behavior, unless $\xi\ll 1$. We illustrate this in Fig.~\ref{fig:CorFRIOlargeN} considering the FRIO discrimination of $N=50$ symmetric states of a qutrit for $\xi=0.01$ and $0.6$. The explanation for this behavior is the negligible effect that separation will have in increasing the distinguishability of the input states which, from the start, are already very poorly distinguishable if $N\gg n$. Hence, the major coherence cost will come from the ME discrimination of the successfully separated states weighted by $P$.

\begin{figure*}[t]
\centerline{\includegraphics[width=\textwidth]{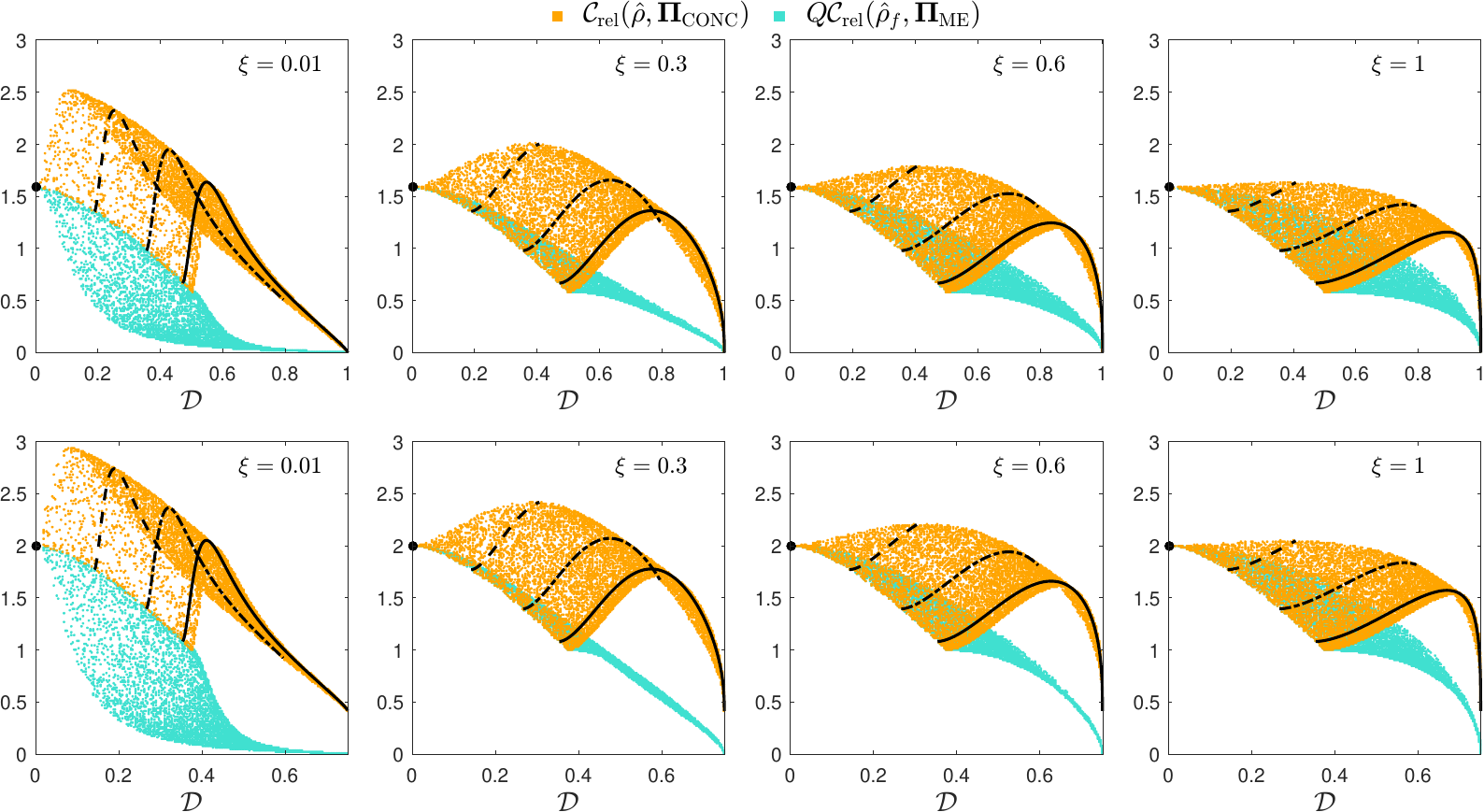}} 
  \caption{POVM coherence (orange dots) and its component $Q\mathcal{C}_\textrm{rel}(\hat{\rho}_f,\bm{\Pi}_\textsc{me})$ (blue dots) vs distinguishability in the concatenated FRIO discrimination of $N=3$ (top row) and $4$ (bottom row) symmetric states of a qutrit. These quantities were computed for $10^4$ random input states and a fixed value of the separation parameter $\xi$ shown in the insets. We also plot the former for inputs in the range $a_\textrm{min}\in[0,a_0]$ [see Eq.~(\ref{eq:coefficients})], where $a_0=0$ (circle), $a_0=0.192$ (dashed lines), $a_0=0.385$ (dash-dotted lines), and $a_0=1/\sqrt{3}$ (solid lines).}
  \label{fig:POVM_CONC}
\end{figure*}


\subsection{POVM coherence in concatenated FRIO measurement} \label{sec:concatenated}

The concatenation of optimal FRIO and ME measurements results in a $2N$-outcome POVM built by the detection operators of Eq.~(\ref{eq:POVM_conc}); the associated probabilities $\{p_j,p_j^?\}_{j=0}^{N-1}$ will be given by 
\begin{subequations}
\begin{align}
    p_j &= \mathrm{Tr} [\hat{A}_j\hat{\rho}\hat{A}_j^\dag] = \frac{P}{N},  \label{eq:pjconc} \\
    p^?_j &= \mathrm{Tr} [\hat{A}_j^?\hat{\rho}\hat{A}_j^{?\dag}] = \frac{Q}{N}.  \label{eq:pfconc}
\end{align}
\end{subequations}
In this case, the postmeasurement states from both conclusive and inconclusive outcomes are pure, so from Eq.~(\ref{eq:POVMcoherence}) the required coherence to implement this POVM on $\hat{\rho}$ will be
\begin{align}
   \mathcal{C}_\textrm{rel}(\hat{\rho},\mathbf{\Pi}_{\textsc{conc}}) = H_2(P) +\log_2 N-S(\hat{\rho}). \label{eq:concatenatedcoherence}
  \end{align}
After simple algebraic manipulation, we can rewrite this expression as 
\begin{align}
    \mathcal{C}_\textrm{rel}(\hat{\rho},\mathbf{\Pi}_{\textsc{conc}})  = \mathcal{C}_\textrm{rel}(\hat{\rho},\mathbf{\Pi}_{\textsc{frio}}) +Q \mathcal{C}_\textrm{rel}(\hat{\rho}_f,\mathbf{\Pi}_{\textsc{me}}), \label{eq:sumofcoherences2}
\end{align}
where $\mathcal{C}_\textrm{rel}(\hat{\rho},\mathbf{\Pi}_{\textsc{frio}})$ is the coherence for the standard FRIO measurement given by Eq.~(\ref{eq:sumofcoherences}) and $\mathcal{C}_\textrm{rel}(\hat{\rho}_f, \mathbf{\Pi}_{\textsc{me}}) = \log_2 N - S(\hat{\rho}_f)$ is the coherence required for discriminating the failure outputs of state separation via ME measurement [see Eq.~(\ref{eq:MEcoherence})]. This latter coherence is bounded by $\log_2N/(n-\mu)\leqslant\mathcal{C}_\textrm{rel}(\hat{\rho}_f, \mathbf{\Pi}_{\textsc{me}})\leqslant \log_2N$, where the lower bound is attained for a maximally mixed $\hat{\rho}_f$ in an $(n-\mu)$-dimensional space, while the upper bound is achieved for a pure $\hat{\rho}_f$. Equation~(\ref{eq:sumofcoherences2}) shows that the POVM coherence in the concatenated measurement satisfies $\mathcal{C}_\textrm{rel}(\hat{\rho},\mathbf{\Pi}_{\textsc{conc}}) \geqslant \mathcal{C}_\textrm{rel}(\hat{\rho},\mathbf{\Pi}_{\textsc{frio}})$, with equality holding only for $Q=0$, which occurs when the input states are maximally distinguishable or $\xi=0$ (when both strategies reduce to the ME measurement).


\subsubsection{Example: Discriminating \texorpdfstring{$N$}{N} symmetric states of a qutrit with concatenated FRIO}

Here, we conclude the example discussed in the previous subsections, now studying the coherence in the concatenated FRIO discrimination between $N$ symmetric states of a qutrit. Once again, we resort to $10^4$ random input states and the fixed inputs defined by Eq.~(\ref{eq:coefficients}). Using Eq.~(\ref{eq:sumofcoherences2}), in Fig.~\ref{fig:POVM_CONC} we plot $\mathcal{C}_\textrm{rel}(\hat{\rho},\mathbf{\Pi}_{\textsc{conc}})$ for both random (orange dots) and fixed inputs (black circle and lines), as well  as the second term on the right-hand side (blue dots) as a function of $\mathcal{D}$ and $\xi$; the top (bottom) row corresponds to $N=3$ ($N=4$) states to be discriminated. The contribution from the first term on the right-hand side of (\ref{eq:sumofcoherences2}), namely the POVM coherence for standard FRIO, was  studied earlier and is shown in Fig.~\ref{fig:POVM_FRIO}. 

The results in Fig.~\ref{fig:POVM_CONC} show that, in regard to the POVM coherence, the concatenated strategy presents similar aspects of both ME and standard FRIO measurements. Like ME, if the input states are parallel, i.e. $\mathcal{D}=0$, we have $\mathcal{C}_\textrm{rel}(\hat{\rho},\mathbf{\Pi}_{\textsc{conc}}) = \mathcal{C}_\textrm{rel}(\hat{\rho},\mathbf{\Pi}_{\textsc{me}})=\log_2N$ for any $\xi$, as a consequence that a ME measurement is implemented on the failure outputs. On the other hand, like the standard FRIO, the POVM coherence presents a monotonic increasing or nonmonotonic behavior with respect to $\mathcal{D}$, as shown in the plots for the fixed inputs. This is a feature inherited from the state separation stage.

The term $Q\mathcal{C}_\textrm{rel} (\hat{\rho}_f,\hat{\mathbf{\Pi}}_\textsc{me})$ in Eq.~(\ref{eq:sumofcoherences2}) represents the extra amount of coherence that needs to be consumed for implementing the concatenated strategy instead of the standard one. It decreases with the distinguishability and increases with the separation degree. This latter behavior is dictated only by the failure rate $Q$, since $\mathcal{C}_\textrm{rel} (\hat{\rho}_f,\hat{\mathbf{\Pi}}_\textsc{me})$ does not depend on $\xi$. We also note from Fig.~\ref{fig:POVM_CONC} that  $\mathcal{C}_\textrm{rel}(\hat{\rho},\hat{\mathbf{\Pi}}_\textsc{conc}) \approx Q \mathcal{C}_\textrm{rel}(\hat{\rho}_f,\hat{\mathbf{\Pi}}_\textsc{me})$ for states which $\mathcal{C}_\textrm{rel}(\hat{\rho},\hat{\mathbf{\Pi}}_\textsc{frio}) \approx 0$, namely the states with very small $a_\textrm{min}$, for which $P\approx 0$ and $S(\hat{\rho}_f)\approx S(\hat{\rho})$ for any $\xi$.


\section{Discussion}
\label{sec:discussion}

\subsection{Bounds of POVM-based coherence}
\label{subsec:Bounds}

In Ref.~\cite{Bischof19}, Bischof \textit{et al.}\ show that for an $N'$-outcome POVM $\bm{\Pi}'=\{\hat{\Pi}'_i\}_{i=0}^{N'-1}$, the relative entropy of POVM-based coherence is bounded by $0\leqslant\mathcal{C}_\textrm{rel}(\hat{\rho},\bm{\Pi}')\leqslant \log_2N'$. The upper bound is attained by the pure states that generate the highest entropy of measurement outcomes. The lower bound is attained for states that satisfy $\hat{\Pi}'_i\hat{\rho}\hat{\Pi}'_j=0$ for all $i\neq j$, and $\hat{\rho}$ is POVM incoherent. 

For the ME POVM there are $N$ pure states whose coherence reaches the upper bound $\log_2 N$. They are given by $|j\rangle=\frac{1}{\sqrt{N}}\sum_k\omega^{-jk}|u_k\rangle$ for $j=0,\ldots,N-1$, where $\{|u_k\rangle\}$ is the orthonormal basis given by the uniform states of Eq.~(\ref{eq:uj}) with $y_k=1$ $\forall k$.  A ME measurement [see Eq.~(\ref{eq:MEPOVM})] on $|j\rangle$ yields, randomly, any outcome $k$ with probability $1/N$, generating the maximal randomness. On the other hand, $\hat{\rho}=\frac{1}{n}\sum_{k=0}^{N-1}y_k|k\rangle\langle k|$ is the only state with minimum coherence, $\log_2N/n$. If $N=n$, this is the maximally mixed state $\hat{\rho}=\hat{I}/n$, which is then POVM incoherent. These bounds can be visualized in the example of Fig.~\ref{fig:POVM_ME} and extend the findings of Ref.~\cite{Bischof19} that considered the ME discrimination of symmetric states for $N=3$ and $n=2$.

For the standard and concatenated FRIO POVMs the upper bounds of coherence are $\log_2(N+1)$ and $\log_22N$, respectively. However, in both cases, there are no states that reach these bounds, no matter the value of $\xi$. The lower bound of coherence is attained both by $\{|j\rangle\}_{j=0}^{N-1}$ and $\hat{\rho}=\hat{I}/n$ in the standard case, and only by the latter in the concatenated case. The results shown in the examples of Figs.~\ref{fig:POVM_FRIO}, \ref{fig:CorFRIOlargeN}, and \ref{fig:POVM_CONC} illustrate this discussion.

\subsection{Private randomness}
\label{subsec:Private}

Quantum coherence is an operationally relevant quantity for quantum cryptography as it quantifies the private randomness of a measurement with respect to eavesdropping activities \cite{Yuan15, Yuan19,Bischof21}, an important result to secure quantum generation of random numbers \cite{Ma19, Ma19-2}. Briefly, let $\hat{\rho}_a$ be the state to be measured and consider an eavesdropper $e$ that has access to a purification $\hat{\rho}_{ae}= |\psi\rangle\langle\psi|_{ae}$, so that $\hat{\rho}_a= \mathrm{Tr}_e \hat{\rho}_{ae}$. Applying the POVM $\mathbf{\Pi}= \lbrace \hat{\Pi}_i = \hat{A}_i^\dagger \hat{A}_i\rbrace$ on $\hat{\rho}_a$ and storing the outcomes $i$ in a register $x$ produces the output joint state $\hat{\rho}'_{{xae}}=\sum_i p_i |i\rangle\langle i|_x\otimes |\psi_i\rangle\langle\psi_i|_{ae}$, where $|\psi_i\rangle_{ae}=\frac{1}{\sqrt{p_i}}(\hat{A}_i\otimes \hat{I})|\psi\rangle_{ae}$ and $p_i = \textrm{Tr} (\hat{\Pi}_i \hat{\rho}_a)$. Bischof {\it et al.}\ \cite{Bischof21} define the randomness of the measurement of $\mathbf{\Pi}$ as $R_{x|e}(\hat{\rho}_a)\equiv\min_{\hat{\rho}_{ae}} S(x|e)_{\hat{\rho}'_{xe}}$, where $S(x|e)_{\hat{\rho}'_{xe}} =S(\hat{\rho}'_{xe})-S(\hat{\rho}'_e)$ denotes the conditional von Neumann entropy, $\hat{\rho}'_{xe}=\textrm{Tr}_a\hat{\rho}'_{{xae}}$,  $\hat{\rho}'_e=\textrm{Tr}_{xa}\hat{\rho}'_{{xae}}$, and the minimization is taken over all purifications. Then they show that
\begin{align}
R_{x|e}(\hat{\rho}_a) = \mathcal{C}_\textrm{rel} (\hat{\rho}_a, \mathbf{\Pi}),
\end{align}
i.e., the POVM coherence quantifies the rate of measurement outcomes that are unpredictable to the eavesdropper. This is a generalization of previous results by Yuan {\it et al.}\ \cite{Yuan15, Yuan19} concerning the standard relative entropy of coherence.

In the state discrimination scenario, if the inputs are orthogonal, then $\hat{\rho}_a= \hat{I}/N$, and the eavesdropper will hold a maximally entangled purification $|\psi\rangle_{ae}=\frac{1}{\sqrt{N}}\sum_{j=0}^{N-1} |u_j\rangle_a \otimes|j\rangle_e$. In this case, e can always uncover the bits generated by the measurement, making secrecy unreachable. Therefore, the measurement outcome privacy lies on the nonorthogonality of the input states, which is verified by the fact that coherence vanishes in all cases where the inputs are perfectly distinguishable (e.g., see Figs.~\ref{fig:POVM_ME}, \ref{fig:POVM_FRIO}, and \ref{fig:POVM_CONC}). 

Our results show that there are optimal intermediate values for the input set distinguishability in terms of maximizing coherence (see Figs.~\ref{fig:POVM_FRIO} and \ref{fig:POVM_CONC}). This feature is not observed for the ME measurement, for which coherence decreases monotonically with distinguishability (Fig.~\ref{fig:POVM_ME}): this means that the maximum coherence is achieved for identical states, for which the discrimination is useless. In this way, the standard and concatenated FRIO schemes benefit scenarios where both the ability to discriminate quantum states and the outcomes secrecy with respect to eavesdropping are relevant figures of merit. In both cases, the optimal distinguishability that maximizes coherence increases with the separation parameter $\xi$, a feature that is more pronounced for the standard measurement. Note that for any $\xi$ the concatenated strategy outperforms the standard one (including ME) in terms of private bit generation. Take for example the case of $N=4$ states of a qutrit and separation parameter $\xi=1$ (bottom rows of Figs.~\ref{fig:POVM_FRIO} and \ref{fig:POVM_CONC}): whereas the concatenated scheme can achieve up to two secret bits per measurement the standard strategy only achieves around one bit. 


\section{\label{sec:conclusion}Conclusions and perspectives}

We presented a detailed study on POVM-based coherence as a resource for standard and concatenated FRIO discrimination of equally likely symmetric states in arbitrary dimensions. As a first result, we showed that the POVM coherence in the assisted separation stage decomposes into the coherence of the ancillary state and the quantum discord between the system and the ancilla, evidencing coherence as a more elementary resource than quantum correlations. This relation may pave the way for better understanding of the measure in Eq.~(\ref{eq:POVMcoherence}) from a fundamental perspective. As quantum coherence was traditionally seen as a quantification of the strength of superposition \cite{Baumgratz14, Streltsov17}, and further as a measure of intrinsic randomness \cite{Yuan15, Yuan19, Bischof21}, it is quite relevant that, at least in the particular case addressed here, its formal extension to generalized measurements \cite{Bischof19} embraces the definition of quantum discord at its core. A more general result in this direction was not the purpose of this paper and we leave further investigation on the subject to future research. We also demonstrated that the POVM coherence for standard and concatenated FRIO decomposes into the POVM coherence measures for state separation and ME measurement, weighted by the probabilities of occurrence of each event. We discussed the operational meaning of such resource in terms of private random bit generation and showed how standard and concatenated FRIO measurements can outperform the ME strategy with respect to this figure of merit. In particular, we concluded that by concatenating the failure outputs of state separation we always achieve better secret bit generation rates compared to standard FRIO. These findings may be useful in high-dimensional quantum random number generation and quantum cryptography \cite{Brask17, Yuan15, Yuan19, Bischof21, Ma19, Ma19-2}.

\begin{acknowledgments}
This work was supported by CNPq INCT-IQ (465469/2014-0), CNPq (422300/2021-7) and (303212/2022-5). L. F. M. acknowledges financial support from CAPES - Finance Code 001, and from the project ``Comparative Analysis of P$\&$M Protocols for Quantum Cryptography'' supported by QuIIN - Quantum Industrial Innovation, EMBRAPII CIMATEC Competence Center in Quantum Technologies, with financial resources from the PPI IoT/Manufatura 4.0 / PPI HardwareBR of the MCTI grant number 053/2023 signed with EMBRAPII. O. J. was supported by an internal grant from Universidad Mayor (PUENTE-2024-17).
\end{acknowledgments}


\input{REFS.bbl}

\end{document}

%% file: REFS.bbl
%